\documentclass[ste,twoside]{stefano}

\usepackage{amssymb}
\usepackage{amsmath}
\usepackage{shadow}
\usepackage{fancybox}

\def\0{\mbox{\tiny $0$}}
\def\1{\mbox{\tiny $1$}}
\def\2{\mbox{\tiny $2$}}
\def\3{\mbox{\tiny $3$}}
\def\4{\mbox{\tiny $4$}}
\def\5{\mbox{\tiny $5$}}
\def\6{\mbox{\tiny $6$}}
\def\7{\mbox{\tiny $7$}}
\def\8{\mbox{\tiny $8$}}
\def\9{\mbox{\tiny $9$}}

\def\inv{\mbox{\tiny $-$$1$}}

\def\R{\mathbb{R}}
\def\C{\mathbb{C}}
\def\H{\mathbb{H}}

\def\Ri{\mbox{\tiny $R$}}
\def\Le{\mbox{\tiny $L$}}

\def\c{\mbox{\tiny $c$}}

\def\q{\mbox{\tiny $q$}}
\def\p{\mbox{\tiny $p$}}

\def\h{\mbox{\tiny $h$}}

\def\m{\mbox{\tiny $m$}}

\def\u{\mbox{\tiny $u$}}
\def\v{\mbox{\tiny $v$}}

\def\qm{\mbox{\tiny $-$$q$}}
\def\qa{\mbox{\tiny $a$$-$$q$}}

\def\ij2{\mbox{\tiny $\frac{i+j}{2}$}}
\def\ji2{\mbox{\tiny $\frac{j-i}{2}$}}

\def\i{\mbox{\tiny $i$}}
\def\j{\mbox{\tiny $j$}}

\def\mi{\mbox{\tiny $-$}}
\def\pl{\mbox{\tiny $+$}}

\def\makeheadbox{{%
\hbox to0pt{\vbox{\baselineskip=10dd\hrule\hbox
to\hsize{\vrule\kern3pt\vbox{\kern3pt \hbox{  {\sc solving simple
quaternionic differential equations} } \hbox{ {\sl Journal of
Mathematical Physics {\bf 44}, 2224-2233 (2003) } \hspace*{5.4cm}
$\boldsymbol{\Sigma \delta \Lambda}$ }
\kern3pt}\hfil\kern3pt\vrule}\hrule}%
\hss}}}

\begin{document}

\title{{\large SOLVING SIMPLE QUATERNIONIC DIFFERENTIAL EQUATIONS}}

\author{
Stefano De Leo\inst{1}
\and Gisele C. Ducati\inst{2}
}

\institute{
Department of Applied Mathematics, State University of Campinas\\
PO Box 6065, SP 13083-970, Campinas, Brazil\\
{\em deleo@ime.unicamp.br}
\and
Department of Mathematics, University of Parana\\
PO Box 19081, PR 81531-970, Curitiba, Brazil\\
{\em ducati@mat.ufpr.br}
}


\date{{\em June, 2003}}

\abstract{The renewed interest in investigating quaternionic
quantum mechanics, in particular tunneling effects~\cite{DDN02},
and the recent results on quaternionic differential
operators~\cite{DD01} motivate the study of resolution methods for
quaternionic differential equations. In this paper, by using the
real matrix representation of left/right acting quaternionic
operators, we prove existence and uniqueness for quaternionic
initial value problems,
 discuss  the reduction of order for quaternionic homogeneous differential
equations and extend to the non-commutative case the method of
variation  of parameters. We also show that the standard Wronskian
cannot uniquely be extended to the quaternionic case.
Nevertheless, the {\em absolute value} of the complex Wronskian
admits a {\em non-commutative} extension  for quaternionic
functions of one real variable. Linear dependence and independence
of solutions of homogeneous (right) $\H$-linear differential
equations is then related to this {\em new} functional. Our
discussion is, for simplicity, presented for quaternionic second
order differential equations. This involves no loss of generality.
Definitions and results can be readily extended to the $n$-order
case.}









\titlerunning{QUATERNIONIC DIFFERENTIAL EQUATIONS}

\maketitle


\section*{{\normalsize I. INTRODUCTION}}

Let  $\mathbb{R}$, $\mathbb{C}  \equiv \mbox{span} \, \left\{1,i
\right\}$, and  $\mathbb{H} \equiv \mbox{span} \, \left\{1,i,j,k
\right\}$ be the real, complex , and quaternionic
field~\cite{HAM},
\[
i^{\2}=j^{\2}=k^{\2}=ijk=-1~,\]
  and \[ \mathcal{F} \, : \, \R \to \R \] be
 the set of real functions of real
variable. Through the paper, quaternionic functions of real
variable, $\Psi(x)  \in \H \, \otimes \mathcal{F}$, will be
denoted by Greek letter and constant quaternionic coefficients by
Roman letter. To shorten notation the prime and double prime in
the quaternionic functions shall respectively indicate the first
and second derivative of quaternionic functions with respect to
the real variable $x$,
\[
\Psi' :=  \frac{\mbox{d} \Psi}{\mbox{d}x} ~~~\mbox{and}~~~\Psi''
:= \, \frac{\mbox{d}^{\2} \Psi}{\mbox{d}x^{\2}}~.
\]
Due to the non-commutative nature of quaternions, it is convenient
to distinguish between the left and right action of the quaternionic
imaginary units $i$, $j$, and $k$  by introducing the operators
$L_{\q}$ and $R_{\p}$ whose
action on quaternionic functions $\Psi$ is given by
\begin{equation}
L_{\q} \Psi = q \, \Psi~~~\mbox{and}~~~ R_{\p} \Psi =\Psi \, p~,
\end{equation}

These (left/right acting) quaternionic operators satisfy
\begin{equation}
L_{\q} \, L_{\p} = L_{\q \p}~,~~~R_{\q} \, R_{\p} = R_{\p
\q}~~~\mbox{and}~~~\left[ \, L_{\q} \, , \, R_{\p} \, \right] =
0~,
\end{equation}
and admit for
\[
q=q_{\0}+i \, q_{\1}+j \, q_{\2}+k \, q_{\3}~,~~
p=p_{\0}+i \, p_{\1}+j \, p_{\2}+k \, p_{\3}~,~~
\Psi= \Psi_{\0}+i \, \Psi_{\1}+j \, \Psi_{\2}+k \, \Psi_{\3}~,
\]
the following {\em real} matrix
representation~\cite{DD99,DS00,DSS02}
\begin{equation}
\label{rrep} L_{\q} \leftrightarrow
 \left( \begin{array}{rrrr} q_{\0} & $-$
q_{\1} & $-$
q_{\2} & $-$ q_{\3} \\
q_{\1} & q_{\0} & $-$
q_{\3} & q_{\2} \\
q_{\2} & q_{\3} &
q_{\0} & $-$ q_{\1} \\
q_{\3} & $-$ q_{\2} & q_{\1} &  q_{\0} \end{array}
\right)~,~~~R_{\p} \leftrightarrow  \left(
\begin{array}{rrrr} p_{\0} & $-$ p_{\1} & $-$
p_{\2} & $-$ p_{\3} \\
p_{\1} & p_{\0} &
p_{\3} & $-$ p_{\2} \\
p_{\2} & $-$ p_{\3} &
p_{\0} &  p_{\1} \\
p_{\3} &  p_{\2} & $-$ p_{\1} &  p_{\0} \end{array} \right)~,~~~
\Psi \leftrightarrow
\left[ \begin{array}{c} \Psi_{\0} \\  \Psi_{\1} \\
\Psi_{\2} \\  \Psi_{\3} \end{array} \right] \in \R^{\4} \otimes
\mathcal{F}~.
\end{equation}

\section*{{\normalsize II. EXISTENCE AND UNIQUENESS.}}
In this section we discuss existence and uniqueness for the
quaternionic initial value problem
\begin{equation}
\label{p1}
 \Psi''=\alpha \, \Psi' + \beta \, \Psi +
\rho~,~~\Psi(x_{\0})=f~,~~\Psi'(x_{\0})=g~,
\end{equation}
with $\alpha(x), \, \beta(x) , \,   \rho(x) \in \H \otimes \mathcal{F}$,
$x_{\0} \in I:(x_{\mi},x_{\pl})$ and $f,g \in \H$.\\

\noindent {\underline {\sf Theorem 1}}. Let $\alpha$,
$\beta$ and $\rho$ in Eq.(\ref{p1}) be continuous functions
of $x$ on an open interval $\mbox{I}$ containing the point
$x=x_{\0}$. Then, the initial value problem (\ref{p1}) has a
solution $\Psi$ on this interval and this solution is unique.\\

\noindent {\sf Proof}. By using the real matrix representation
(\ref{rrep}), we can immediately rewrite the quaternionic initial
value problem (\ref{p1}) in the following vector form
\begin{equation}
\label{syst}
\left[ \begin{array}{c} \Psi_{\0} \\  \Psi_{\1} \\
\Psi_{\2} \\  \Psi_{\3} \end{array} \right]'' = \left(
\begin{array}{rrrr} \alpha_{\0} & $-$ \alpha_{\1} & $-$
\alpha_{\2} & $-$ \alpha_{\3} \\
\alpha_{\1} & \alpha_{\0} & $-$
\alpha_{\3} & \alpha_{\2} \\
\alpha_{\2} & \alpha_{\3} &
\alpha_{\0} & $-$ \alpha_{\1} \\
\alpha_{\3} & $-$ \alpha_{\2} & \alpha_{\1} &  \alpha_{\0}
\end{array} \right) \,
\left[ \begin{array}{c} \Psi_{\0} \\  \Psi_{\1} \\
\Psi_{\2} \\  \Psi_{\3} \end{array} \right]'+ \left(
\begin{array}{rrrr} \beta_{\0} & $-$ \beta_{\1} & $-$
\beta_{\2} & $-$ \beta_{\3} \\
\beta_{\1} & \beta_{\0} & $-$
\beta_{\3} & \beta_{\2} \\
\beta_{\2} & \beta_{\3} &
\beta_{\0} & $-$ \beta_{\1} \\
\beta_{\3} & $-$ \beta_{\2} & \beta_{\1} &  \beta_{\0}
\end{array} \right) \,
\left[ \begin{array}{c} \Psi_{\0} \\  \Psi_{\1} \\
\Psi_{\2} \\  \Psi_{\3} \end{array} \right]+
\left[ \begin{array}{c} \rho_{\0} \\  \rho_{\1} \\
\rho_{\2} \\  \rho_{\3} \end{array} \right]
\end{equation}
with
\begin{equation}
\label{icon}
\left[ \begin{array}{c} \Psi_{\0}(x_{\0}) \\  \Psi_{\1}(x_{\0})  \\
\Psi_{\2}(x_{\0})  \\  \Psi_{\3}(x_{\0}) \end{array} \right]
 =
\left[ \begin{array}{c} f_{\0} \\  f_{\1} \\
f_{\2} \\  f_{\3} \end{array} \right]~~~\mbox{and}~~~
\left[ \begin{array}{c} \Psi_{\0}'(x_{\0}) \\  \Psi_{\1}'(x_{\0}) \\
\Psi_{\2}(x_{\0}) \\  \Psi_{\3}'(x_{\0}) \end{array} \right]
=
\left[ \begin{array}{c} g_{\0} \\  g_{\1} \\
g_{\2} \\  g_{\3} \end{array} \right]~.
\end{equation}
Eq.(\ref{syst}) represents a (nonhomogeneous) linear system with
$\alpha_{\m}, \beta_{\m}, \rho_{\m} \in \R \otimes \mathcal{F}$,
where {\small $m=0,1,2,3$}. These functions are (see hypothesis of
Theorem 1) continuous (real) functions of $x$ on an open interval
$I$ containing the point $x=x_{\0}$. Then, by a well-known theorem
of analysis, see for example ref.~\cite{KRE}, the linear system
(\ref{syst}) has a solution
\[
\left[ \begin{array}{c} \Psi_{\0} \\  \Psi_{\1} \\
\Psi_{\2} \\  \Psi_{\3} \end{array} \right] \in \R^{\4} \otimes
\mathcal{F}
\]
on this interval satisfying (\ref{icon}), and this solution is
unique \hfill {\tiny $\blacksquare$}

\section*{{\normalsize III. LINEAR INDEPENDENCE AND DEPENDENCE OF SOLUTIONS}}

Let us now analyse the linear independence and dependence
of the solutions of second order homogeneous differential
equations
\begin{equation}
\label{p1b}
 \Psi''=\alpha \, \Psi' + \beta \, \Psi~,
\end{equation}
where $\alpha$ and $\beta$ are  (quaternionic) continuous
functions of $x$ on an open interval I. Eq.(\ref{p1b}) is linear
over $\H$ from the right. Consequently, if $\varphi$ is a solution
of Eq.(\ref{p1b}) only the function obtained by right
multiplication by constant quaternionic coefficients, $\varphi \,
u$, still represent a solution of such an equation.
 The general solution of Eq.(\ref{p1b}) is
given in terms of a pair of linearly independent solutions
$\varphi$ and $\xi$ by
\begin{equation}
\label{ges} \Psi = \varphi \, u + \xi \, v~,
\end{equation}
where  $\varphi=\varphi_{\0} + i \, \varphi_{\1} + j \,
\varphi_{\2} +k \, \varphi_{\3}$, $\xi=\xi_{\0} + i \, \xi_{\1} +
j \, \xi_{\2} +k \, \xi_{\3}$ $\in \H \, \otimes \mathcal{F}$ and $u,
\, v \in \H$.

In the standard complex theory ($\varphi=\varphi_{\0} + i \,
\varphi_{\1}$ and $\xi=\xi_{\0} + i \, \xi_{\1}$ $\in \C \otimes
\mathcal{F}$)
 a useful criterion to establish
linear independence and dependence of two solutions of homogeneous
second order differential equation,  uses the concept of {\em
Wronskian} of these solutions defined by
\begin{equation}
\label{cw} W = \varphi \, \xi' - \varphi' \, \xi~,~~~ W  \in \C
\otimes \mathcal{F}~.
\end{equation}
This definition cannot be extended to quaternionic functions. Let
us consider two linearly dependent solutions of Eq.(\ref{p1b}),
\begin{equation}
\label{ldc1} \xi = \varphi \, q~,~~~
\varphi, \, \xi \in \H \otimes \mathcal{F}~,~~q\in \H~.
\end{equation}
By substituting $\xi = \varphi \, q$  and
$\xi' = \varphi' \, q$ in the {\em Wronskian} (\ref{cw}), we find
\[
 \varphi \, \xi' - \varphi' \,  \xi =
\varphi  \, \varphi' \, q - \varphi' \, \varphi \, q   \neq  0~.
\]
Observe that a quaternionic function and its first derivative do
not, in general, commute. Thus, the definition (\ref{cw}), and all
its possible factor combinations  cannot be extended to the quaternionic
case.

Let us now use the linear dependence condition (\ref{ldc1}) to
investigate the possibility to define a quaternionic functional
which extends (in a nontrivial way) the standard (complex) {\em
Wronskian} to the noncommutative case. From Eq.(\ref{ldc1})
and its derivative, we
get
\[
q = \varphi^{\inv}   \xi  = \left( \varphi' \right)^{\inv}
\xi'~,\]
where $\varphi^{\inv} \equiv 1/ \varphi$ and $\left(
\varphi' \right)^{\inv} \equiv 1 / \varphi'$.
Consequently, for
linearly dependent quaternionic solutions , we have
\begin{equation}
\label{qw1} \xi' - \varphi' \, \varphi^{\inv} \,  \xi = 0~.
\end{equation}
To recover, in the complex limit, the standard definition
(\ref{cw}) we multiply $\xi' - \varphi' \varphi^{\inv} \xi$ by
$\varphi$. Due to the noncommutative nature of quaternions, we
have to consider the following possibilities
\begin{equation}
\label{w1}
 W_{\Le}  =  \varphi \, \left( \, \xi' - \varphi' \,
\varphi^{\inv} \xi \, \right)~~~\mbox{and}~~~W_{\Ri}  = \left( \,
\xi' - \varphi' \, \varphi^{\inv} \xi \, \right) \, \varphi~.
\end{equation}
Obviously two other {\em similar}  definitions can be obtained
by $\varphi \leftrightarrow \xi$,
\begin{equation}
\label{w2}
\widetilde{W}_{\Le}  =  - \xi \, \left( \, \varphi' - \xi' \,
\xi^{\inv} \varphi \, \right) = - W_{\Le}[\varphi \leftrightarrow
\xi]  ~~~\mbox{and}~~~ \widetilde{W}_{\Ri} = - \left( \, \varphi '
- \xi' \, \xi^{\inv} \varphi \, \right) \, \xi = - W_{\Ri}[\varphi
\leftrightarrow \xi] ~.
\end{equation}
The quaternionic functionals (\ref{w1}) and (\ref{w2}), which give
in the complex limit the standard definition, extend a first
important property of {\em
Wronskian}. Two solutions of Eq.(\ref{p1b}) are linearly
dependent on I
\underline{if} $W_{\Le ( \Ri )}$ $\left[ \widetilde{W}_{\Le ( \Ri)}
\right]$ is zero on I. To
avoid ambiguity in defining the {\em Wronskian}, we shall
introduce a (real) functional,
\[ |W|^{\2}=|W_{\Le}|^{\2}= |W_{\Ri}|^{\2} = | \widetilde{W}_{\Le} |^{\2} = |
\widetilde{W}_{\Ri} |^{\2}~,\]
which extends  the {\em squared absolute value} of the {\em Wronskian}.
This {\em unique} functional is
\begin{equation}
\label{avw} |W|^{\2}  = |\varphi|^{\2} \, |\xi'|^{\2} + |\xi \,
|^{\2} \, |\varphi'|^{\2} - \varphi' \, \varphi_{\c}  \, \xi  \,
\xi'_{\c} - \xi' \, \xi_{\c} \, \varphi \, \varphi'_{\c}~~~\in \R
\otimes \mathcal{F}~,
\end{equation}
where $\varphi_{\c}=\varphi_{\0} - i \, \varphi_{\1} - j \,
\varphi_{\2} - k \, \varphi_{\3}$ and  $\xi_{\c} =\xi_{\0} - i \,
\xi_{\1} - j \, \xi_{\2}  - k \, \xi_{\3}$ are, respectively,  the
quaternionic conjugate functions of $\varphi$ and $\xi$.

Observe that Eq.(\ref{avw}) can also be obtained as an application
of the Dieudonn\'e theory of quaternionic
determinants~\cite{det1,det3,det4,det5,DEL00}. In fact,
\begin{equation}
\label{referee} |W|^{\2} = \left[ \mbox{Det}(M) \right]^{\2} :=
\mbox{det} \left(M \, M^+\right)~,
\end{equation}
where
\[ M = \left( \begin{array}{cc}
\varphi & \xi  \\
\varphi' & \xi'
\end{array}
\right)~.
\]

\vspace*{0.5cm}

\noindent {\underline {\sf Theorem 2}}. Let $\alpha$ and
$\beta$ in Eq.(\ref{p1b}) be continuous functions of $x$ on an
open interval $I\,:\,(a,b)$. Then, two solutions $\varphi$ and
$\xi$ of Eq.(\ref{p1b}) on I are linearly dependent on I if and
only if the {\em absolute value} of the Wronskian, $|W|$, is
zero at some $x_{\0}$ in I.\\

\noindent The proof will be divided into three
steps:\\
{\sf (a)} - If $\varphi$ and $\xi$ are linearly dependent
on I then $|W|=0$;\\
{\sf (b)} - If $|W|=0$ at some $x_{\0}$ in I then $|W| = 0$ on
I.\\
{\sf (c)} - If $|W|=0$ at some $x_{\0}$ in I then $\varphi$ and
$\xi$ are linearly dependent on I.\\

\noindent {\sf Proof (a)}. If $\varphi$ and $\xi$ are linearly
dependent on I, then Eq.(\ref{ldc1}) holds on
I. From Eq.(\ref{ldc1}), we get
\[
|W|^{\2} = |\varphi|^{\2} \, |\varphi'|^{\2} \, |q|^{\2} +
|\varphi|^{\2} \, |\varphi'|^{\2} \, |q|^{\2} - \varphi' \varphi_{\c}
\varphi \, |q|^{\2} \,  \varphi^{\prime}_{\c} - \varphi' \, |q|^{\2} \,
\varphi_{\c} \varphi \varphi^{\prime}_{\c} =0~,
\]
then $|W|=0$.\\

\noindent {\sf Proof (b)}. Let us consider Eq.(\ref{avw}). By
calculating the first derivative of the left and right side term,
we obtain
\[
\begin{array}{cccl}
2 \, |W| \, |W|' & = &  & \varphi' \varphi_{\c} \xi' \xi^{\prime}_{\c}
+ \varphi \varphi^{\prime}_{\c} \xi' \xi^{\prime}_{\c} + \varphi
\varphi_{\c} \Psi''_{\2} \xi^{\prime}_{\c} + \varphi \varphi_{\c} \xi'
\xi^{\prime  \prime}_{\c} + \\
 & & + & \xi' \xi_{\c} \varphi' \varphi^{\prime}_{\c} +
\xi \xi^{\prime}_{\c} \varphi' \varphi^{\prime}_{\c} + \xi \xi_{\c}
\Psi''_{\1} \varphi^{\prime}_{\c} +
\xi \xi_{\c} \varphi' \varphi^{\prime \prime}_{\c} + \\
& & - & \Psi''_{\1} \varphi_{\c} \xi \xi^{\prime}_{\c} - \varphi'
\varphi^{\prime}_{\c} \xi \xi^{\prime}_{\c} - \varphi' \varphi_{\c} \xi'
\xi^{\prime}_{\c} -
\varphi' \varphi_{\c} \xi \xi^{\prime \prime}_{\c} +\\
 & & - &  \Psi''_{\2} \xi_{\c} \varphi \varphi^{\prime}_{\c} -
 \xi' \xi^{\prime}_{\c} \varphi \varphi^{\prime}_{\c} -
 \xi' \xi_{\c} \varphi' \varphi^{\prime}_{\c} -
 \xi' \xi_{\c} \varphi \varphi^{\prime \prime}_{\c}\\
  & = & & |\varphi|^{\2}  ( \Psi''_{\2}
\xi^{\prime}_{\c} + \xi' \xi^{\prime  \prime}_{\c}) + |\xi|^{\2} (
\Psi''_{\1} \varphi^{\prime}_{\c} + \varphi'
\varphi^{\prime  \prime}_{\c})\,  + \\
 & & - & \Psi''_{\1} \varphi_{\c} \xi \xi^{\prime}_{\c} -
 \varphi' \varphi_{\c} \xi \xi^{\prime \prime}_{\c} -
 \Psi''_{\2} \xi_{\c} \varphi \varphi^{\prime}_{\c} -
 \xi' \xi_{\c} \varphi \varphi^{\prime \prime}_{\c}\\
& = & & |\varphi|^{\2}  (\alpha \, |\xi'|^{\2} + \beta \, \xi
\xi^{\prime}_{\c} + \mbox{h.c.} ) + |\xi|^{\2} (\alpha \,
|\varphi'|^{\2} + \beta \, \varphi
\varphi^{\prime}_{\c} + \mbox{h.c.} ) \, + \\
 & & - & [(\alpha \, \varphi' \varphi_{\c} + \beta \, |\varphi|^{\2})
  \, \xi \xi^{\prime}_{\c} + \mbox{h.c.} ] -
[(\alpha \, \xi' \xi_{\c} + \beta \, |\xi|^{\2})
  \, \varphi \varphi^{\prime}_{\c} + \mbox{h.c.} ]\\
 & = & & 2 \, \mbox{Re} [\alpha] \,  ( |\varphi|^{\2} \, |\xi'|^{\2} +
|\xi|^{\2} \, |\varphi'|^{\2} - \varphi' \varphi_{\c} \xi
\xi^{\prime}_{\c} - \xi' \xi_{\c} \varphi
\varphi^{\prime}_{\c})\\
& = & & 2 \, \mbox{Re} [\alpha] \, |W|^{\2}~.
 \end{array}
\]
By a simple integration, we find
\begin{equation}  |W(x)| = \exp
\left[ \int_{x_{\0}}^{x} \mbox{Re}[\alpha(y)]\, \mbox{d} y \right]
\, |W(x_{\0})|~.
\end{equation}
This prove the statement (b).\\

\noindent {\sf Proof (c)}. From the statement (b), we have
\[ |W(x_{\0})|=0~~~\Rightarrow~~~|W(x)|=0~,~~~x\in \mbox{I}~.\]
This implies that the quaternionic matrix
\[ \left( \begin{array}{cc}
\varphi & \xi  \\
\varphi' & \xi'
\end{array}
\right)
\]
is not invertible on I~\cite{DEL00}. Hence the linear system
\begin{eqnarray*}
\varphi  \,  q_{\1} + \xi \,  q_{\2} ~ & = & 0~,\\
\varphi' \,  q_{\1} + \xi' \, q_{\2} & = & 0~,
\end{eqnarray*}
in the unknowns $q_{\1,\2}\in \H$, has a solution
$(q_{\1},q_{\2})$ where $q_{\1}$ and $q_{\2}$ are not both zero.
Recalling that $\varphi$ and $\xi$ are linearly independent on an
interval I if
\[
\varphi(x) \,  q_{\1} + \xi(x) \, q_{\2} =
0~~~\Rightarrow~~~q_{\1}=q_{\2}=0~,
\]
the fact that $q_{\1}$ and $q_{\2}$ are not both zero guarantees
the linear dependence of $\varphi$ and $\xi$ on I
\hfill {\tiny $\blacksquare$}\\

\noindent {\underline {\sf Example 1}}. Show that $\varphi=\exp
[-ix]$ and $\xi=\exp [(i-j)x]$ form a basis of solutions of
\begin{equation}
\label{ex1} \Psi''+j\, \Psi'+(1-k) \, \Psi =0~,
\end{equation}
on any interval.\\

\noindent {\sf Solution}. Substitution shows that they are
solutions,
\begin{eqnarray*}
\left[-1 + j (-i) + 1 - k \right] \, \exp [-ix] & = & 0~,\\
\left[-2 + j (i-j) + 1 - k \right] \, \exp [(i-j)x] & = & 0~,
\end{eqnarray*}
and linear independence follows from Theorem 2, since
\[
|W|= \sqrt{|i-j|^{\2} + |i|^{\2} + i (j-i) - (i-j) i} = \sqrt{5}~.
\]

\section*{{\normalsize IV. HOMOGENEOUS EQUATIONS: REDUCTION OF ORDER}}

Let $\varphi$ be solution of Eq.(\ref{p1b}) on some interval I.
Looking for a solution in the form
\[ \xi = \varphi \, \tau\]
and substituting $\xi$ and its derivatives
\[ \xi' = \varphi' \, \tau +\varphi \,
\tau'~~~\mbox{and}~~~ \xi'' = \varphi'' \, \tau +2 \,
\varphi' \, \tau' + \varphi \, \tau''
\]
into Eq.(\ref{p1b}), we obtain
\begin{equation}
\label{red} \tau'' =  \left( \varphi^{\inv} \alpha \, \varphi -
2 \, \varphi^{\inv} \varphi' \right) \, \tau'~.
\end{equation}
It is important to observe that, for quaternionic functions, we
{\em cannot} give a formal solution of the previous equation. Only
in particular cases, Eq.(\ref{red}) can be immediately integrated.
For example, for homogeneous second order equations with constant
coefficients,
\[ \alpha(x) \to a \in \H~~~\mbox{and}~~~\beta(x) \to b \in \H~,\]
at least one solution is in the form of a quaternionic
exponential, $\varphi=\exp[qx]$, and consequently Eq.(\ref{red})
reduces to
\begin{equation}
\label{red2} \varphi \tau'' =  \left( a - 2 \, q \right) \,
\varphi \tau'~.
\end{equation}
Let us introduce the quaternionic function
\[ \sigma = \varphi \tau'~.\]
Observing that
\[ \sigma'=
\varphi' \tau' +\varphi \tau''= q \, \varphi \tau' + \varphi
\tau''~,\] Eq.(\ref{red2}) can be rewritten as follows
\begin{equation}
\label{red3}
\sigma' =  \left( a - q \right) \sigma~.
\end{equation}
This equation can be immediately integrated, its solution reads
\[
\sigma = \exp[(a-q) x]~.
\]
Thus, the second solution of the homogeneous second order
differential equation with constant coefficients is given by
\begin{equation}
\label{sec} \xi = \exp[qx] \int \exp[-qx] \, \exp[(a-q)x] \,
\mbox{d}x~.
\end{equation}
In the complex limit ($a,q \in \C$) we find the well-known results
$\xi\propto \exp[(a-q)x]$ if $2q\neq a$ and $\xi\propto x
\exp[qx]$ if $2q = a$. In the quaternionic case ($a,q \in \C$),
the integral which appears in (\ref{sec}) must be treated with
care. The solution of this integral will give interesting
information about the second solution of quaternionic
differential equations with constant coefficients when the
associated characteristic quadratic equation has a unique
solution. To solve the integral in Eq.(\ref{sec}), we start by
observing that
\[
[e^{ux} \, e^{vx} ]' = u \, e^{ux} \, e^{vx} + e^{ux} \, e^{vx} \,
v  = \left( L_{\u} + R_{\v} \right) \, e^{ux} \, e^{vx}~.
\]
If the operator $L_{\u} + R_{\v}$  is invertible  the previous
equality implies
\[ \int e^{ux} \, e^{vx} \, \mbox{d}x =
\left( L_{\u} + R_{\v} \right)^{\inv} \, e^{ux} \, e^{vx}~.
\]
This result guarantees that, if the operator $L_{\qm} - R_{\qa}$
is invertible  the second solution can be written in the form
\begin{eqnarray}
\label{sec1} \xi &=& \exp[qx] \, \left( L_{\qm} + R_{\qa}
\right)^{\inv}  \, \exp[-qx] \, \exp[(a-q)x] \nonumber\\
 &=& \exp[qx] \, \left( R_{\qa} - L_{\q} \right)^{\inv} \,  \exp[-qx] \,
\exp[(a-q)x] ~.
\end{eqnarray}
If the operator $L_{\qm} + R_{\qa}$ is {\em not} invertible, we
need to solve the integral which appears in (\ref{sec}) by using
the polar decomposition of quaternions (see example 3) and a term
linearly dependent on $x$ will appear. In the complex case ($a,q\in
\C$), the operator $L_{\qm} + R_{\qa}$ is not invertible if and
only if $2q=a$. In the quaternionic ($a,q\in \H$), the condition
$2q\neq a$ does not guarantee that the operator is invertible.\\

\noindent {\underline {\sf Example 2}}. Knowing that
$\varphi=\exp [-ix]$ is solution of the homogeneous second order
equation (\ref{ex1}), find (by using the method of reduction of
order)
a second independent solution, $\xi$.\\

\noindent {\sf Solution}. We have $q=-i$ and $a=-j$. To use
Eq.(\ref{sec1}) we have to prove that the operator
\[ L_{\qm} + R_{\qa} = L_{\i} + R_{\i\mi\j} \]
is invertible. A simple algebraic calculation shows that
\[\left( L_{\i} - R_{\i\mi\j} \right) \left( L_{\i} + R_{\i\mi\j}
\right) = 1~. \]
Thus,
\[ \left( L_{\qm} + R_{\qa} \right)^{\inv} = L_{\i} - R_{\i\mi\j}~. \]
We are now ready to calculate $\xi$ from Eq.(\ref{sec1}),
\[
\xi = \exp [-ix] \, \left( L_{\i} - R_{\i\mi\j} \right) \, \exp
[ix] \, \exp[(i-j)x] = \left( L_{\i} - R_{\i\mi\j} \right) \,
\exp[(i-j)x] = \exp[(i-j)x] \, j~.
\]
Due to the $\H$ linearity (from the right) of Eq.(\ref{ex1}) the
right factor $j$ can be ignored recovering the solution of example
1.\\

\noindent {\underline {\sf Example 3}}. Inspection shows that
\begin{equation}
\label{ex3} \Psi'' + i \, \Psi'+ \mbox{$\frac{k}{2}$} =0
\end{equation}
has $\varphi=\exp[-\frac{i+j}{2}\, x]$ as a first solution. Find
the second linear independent solution.\\

\noindent {\sf Solution}. We have $q=-\frac{i+j}{2}$ and $a=-i$.
In this case,  the operator
\[ L_{\qm} + R_{\qa} = L_{\ij2} + R_{\ji2} \]
is {\em not} invertible. This is easily seen by using, for
example, the real matrix representation (\ref{rrep}). Thus, the
integral in Eq.(\ref{sec}) cannot be expressed in terms of an
exponential product. Let us explicitly calculate $\xi$ from
Eq.(\ref{sec}). We find
\begin{eqnarray*}
\xi & = & \exp[ - \mbox{$\frac{i+j}{2}$} \, x] \, \int \,
\exp[\mbox{$\frac{i+j}{2}$} \, x] \, \exp[\mbox{$\frac{j-i}{2}$}
\, x] \, \mbox{d}
x\\
 & = &\exp[ - \mbox{$\frac{i+j}{2}$} \, x] \, \int \,
\left( \cos \mbox{$\frac{x}{\sqrt{2}}$} +
\mbox{$\frac{i+j}{\sqrt{2}}$} \, \sin \mbox{$\frac{x}{\sqrt{2}}$}
\right) \, \left( \cos \mbox{$\frac{x}{\sqrt{2}}$} +
\mbox{$\frac{j-i}{\sqrt{2}}$} \, \sin \mbox{$\frac{x}{\sqrt{2}}$}
\right) \, \mbox{d} x\\
 & = &\exp[ - \mbox{$\frac{i+j}{2}$} \, x] \, \int \,
\left\{ 1 - k \, \exp [ - (i+j) x] \right\} \,
\mbox{$\frac{1+k}{2}$} \, \, \mbox{d} x~.
\end{eqnarray*}
Due to the $\H$ linearity (from the right) of Eq.(\ref{ex3}) the
right factor $\frac{1+k}{2}$ can be removed. After integration, we
find
\[
\xi =\exp[ - \mbox{$\frac{i+j}{2}$} \, x] \, \left\{ x -k \,
\mbox{$\frac{i+j}{2}$} \, \exp [ - (i+j) x] \right\} = \left( x +
\mbox{$\frac{i-j}{2}$} \right) \, \exp[ - \mbox{$\frac{i+j}{2}$}
\, x]~.
\]
Observe that the quaternionic factor $\frac{i-j}{2}$ appears on
the left of the quaternionic exponential and consequently {\em
cannot} be removed. It is a fundamental part of the solution.
Inspection shows that
\[ \xi = x \exp[ -
\mbox{$\frac{i+j}{2}$} \, x] \]
is {\em not} solution of
Eq.(\ref{ex3}).

\section*{{\normalsize V. NONHOMOGENEOUS EQUATIONS: VARIATION OF PARAMETERS}}

A general solution of the nonhomogeneous equation (\ref{p1}) is a
solution of the form
\begin{equation}
\label{gsol} \Psi = \Psi_{\h} + \Psi_{\p}~,
\end{equation}
where
\[
\Psi_{\h} = \varphi \, q_{\1} + \xi \, q_{\2} \] is a general
solution of the homogeneous equation (\ref{p1b}) and $\Psi_{\p}$
is any particular solution of (\ref{p1}) containing no arbitrary
constants. In this section we discuss the so-called method of
variation of parameters to find a particular solution for
quaternionic nonhomogeneous differential equations.

A method to solve a homogeneous second order quaternionic
differential equations with constant coefficients has been
recently developed~\cite{DD01}. Quaternionic differential
equations with non constant coefficients are under investigation.
We suppose to know two independent solutions of the homogeneous
equation associated with Eq.(\ref{p1b}). We wish to investigate if
the method of variation of parameters still works in the
quaternionic case.

The method of variation of parameters involves replacing the
constant $q_{\1}$ and $q_{\2}$
by quaternionic functions $\nu_{\1}(x)$ and $\nu_{\2}(x)$
to be determined so that the resulting function
\[ \Psi_{\p} = \varphi \, \nu_{\1} + \xi \, \nu_{\2} \]
is a particular solution of Eq.(\ref{p1}). By differentiating
$\Psi_{\p}$ we obtain
\[ \Psi'_{\p} = \varphi' \nu_{\1} + \xi' \nu_{\2} +
\varphi \, \nu'_{\1} + \xi \,\nu'_{\2}~.\] The requirement that
$\Psi_{\p}$ satisfies Eq.\,(\ref{p1}) imposes only {\em one}
condition on $\nu_{\1}$ and $\nu_{\2}$. Hence, we can impose a
second arbitrary condition, that is
\begin{equation}
\label{cd1}  \varphi \, \nu'_{\1} + \xi \, \nu'_{\2}=0~.
\end{equation}
This reduces $\Psi'_{\p}$ to the form
\[ \Psi'_{\p} = \varphi' \nu_{\1} + \xi' \nu_{\2}~.\]
By differentiating this function we have
\[
\Psi''_{\p} = \varphi'' \nu_{\1} + \varphi' \nu'_{\1} +
\xi'' \nu_{\2} + \xi' \nu'_{\2}~.\]
Substituting
$\Psi_{\p}$, $\Psi'_{\p}$, and $\Psi''_{\p}$ in Eq.(\ref{p1}) we
readily obtain
\begin{equation}
\label{cd2} \varphi' \nu'_{\1} + \xi' \nu'_{\2} =\rho~.
\end{equation}
Collecting Eq.(\ref{cd1}) and Eq.(\ref{cd2}), we can construct the
following matrix system
\begin{equation}
\label{msys} \left( \begin{array}{cc}
          \varphi & \xi\\
          \varphi' & \xi'
       \end{array}  \right) \left[ \begin{array}{c}
                                      \nu'_{\1}\\
                                      \nu'_{\2}
                                   \end{array}  \right] =
\left[ \begin{array}{c}
           0\\
           \rho
       \end{array}  \right]~,
\end{equation}
from which ($|W|\neq 0 $) we obtain
\begin{eqnarray}
\label{msys2} \left[ \begin{array}{c}
                                      \nu'_{\1}\\
                                      \nu'_{\2}
                                   \end{array}  \right] & = &
\left( \begin{array}{cc}
          \varphi & \xi\\
          \varphi' & \xi'
       \end{array}  \right)^{\inv}
\left[ \begin{array}{c}
           0\\
           \rho
       \end{array}  \right] \nonumber \\
 & = &
\left(
\begin{array}{cc}
 \left[ \varphi - \xi \xi^{\prime \inv} \varphi'
\right]^{\inv} & \left[ \varphi' - \xi' \xi^{-1}
\varphi \right]^{\inv}\\
 \left[ \xi - \varphi \varphi^{\prime
\inv} \xi' \right]^{\inv} &
 \left[ \xi' - \varphi' \varphi^{-1} \xi \right]^{\inv}
\end{array}
\right) \left[ \begin{array}{c}
           0\\
           \rho
       \end{array}  \right]
\end{eqnarray}
Then,
\begin{equation}
\label{ieq} \nu'_{\1} =  [  \varphi' - \xi' \xi^{\inv} \varphi
]^{\inv} \rho~~~\mbox{and}~~~
 \nu'_{\2}= \left[ \xi' - \varphi' \varphi^{-1}
 \xi \right]^{\inv} \rho~.
\end{equation}
To find $\nu_{\1}(x)$ and $\nu_{\2}(x)$ we have to integrate the
previous equations.\\

\noindent {\underline {\sf Example 4}}. Find a general solution of
the nonhomogeneous quaternionic differential equation
\begin{equation}
\label{ex4} \Psi'' + j \, \Psi' + (1 - k) \,  \Psi = i \, x
\end{equation}

\noindent {\sf Solution}. The solution of the associated
homogeneous equation (see example 1)  is
\[
\Psi_{\h} = \exp [-i x] \, q_{\1} + \exp [-(i + j)x] \, q_{\2}~.
\]
The particular solution is
\[
\Psi_{\p} = \exp [-i x] \, \nu_{\1} + \exp [-(i + j)x] \,
\nu_{\2}~.
\]
Consequently, from Eqs.(\ref{ieq}) we find
\[
\nu'_{\1}= \exp[ix] \, x \, k~~~\mbox{and}~~~ \nu'_{\2}= -
\exp[(i + j)x] \, x \, k
\]
which after integration give
\[
\nu_{\1}(x) = (1-ix) \, \exp[ix] \, k~~~\mbox{and}~~~
 \nu_{\2}(x) = - \mbox{$\frac{1}{2}$} \,  [1 - (i+j)x] \,
 \exp[(i + j)x] \, k~.
\]
Finally
\[
\Psi_{p} =\mbox{$\frac{1}{2}$} \,  \left[ (i + j) \, x + k
\right]~.
\]
A general solution of Eq.(\ref{ex3}) is
\[
\Psi = \exp [-i x] \, q_{\1} + \exp [-(i + j)x] \, q_{\2} +
\mbox{$\frac{1}{2}$} \,  \left[ (i + j) \, x + k \right]~.
\]

\section*{{\normalsize VI.CONCLUSIONS AND OUTLOOKS}}

The recent results on violations of quantum mechanics by
quaternionic potentials~\cite{DDN02} and the possibility to get a
better understanding of CP-violation phenomena within a
quaternionic formulation of physical theories~\cite{DDN02,ADL}
stimulated the study of quaternionic differential
operators~\cite{DD01}. In this paper, we have proved existence and
uniqueness for quaternionic initial value problems and solved
simple quaternionic differential equations by discussing the
reduction of order for quaternionic homogeneous equations and by
extending to the non-commutative case the method of variation of
parameters and the definition of absolute value of the Wronskian
functional.

In view of a more complete discussion of quantum dynamical systems
using quaternionic wave packets, our next research (mathematical)
interest will be the study of quaternionic integral transforms.
The quaternionic formulation of Fourier transforms could find an
immediate and interesting application in the study of delay time
modifications of wave packets scattered by a quaternionic
potential step.

\section*{{\normalsize Acknowledgements}}

The authors thank the referee for comments, references, and
suggestions, and for drawing their attention to an alternative way
to obtain the squared absolute value of the Wronskian (\ref{avw})
by Diuedonn\'e's theory of determinants [see Eq.(\ref{referee})].

\end{document}